\documentclass[prd,showpacs,showkeys,nofootinbib,floatfix,eqsecnum,
               fleqn,preprint,12pt,tightenlines]{revtex4-1}

\usepackage{amsmath,amssymb,revsymb,graphicx,dcolumn}
\usepackage{array}
\newcommand{\beq}{\begin{equation}}
\newcommand{\eeq}{\end{equation}}
\newcommand{\beqa}{\begin{eqnarray}}
\newcommand{\eeqa}{\end{eqnarray}}
\newcommand{\bsubeqs}{\begin{subequations}}
\newcommand{\esubeqs}{\end{subequations}}

\begin{document}

\begin{widetext}
%\noindent Phys. Rev. D  \hfill  arXiv:1911.06173\;(\version)
\noindent Phys. Rev. D {\bf 101}, 064061 (2020) \hfill  arXiv:1911.06173
%
%\noindent arXiv:1911.06173 \hfill KA--TP--20--2019\;(\version)
\newline\vspace*{5mm}
\end{widetext}

\title{Nonsingular bouncing cosmology from general relativity:\\
Scalar metric perturbations\vspace*{4mm}}

 \author{F.R. Klinkhamer}
 \email{frans.klinkhamer@kit.edu}

 \affiliation{Institute for Theoretical Physics,
 Karlsruhe Institute of Technology (KIT),\\
 76128 Karlsruhe, Germany\\}

 \author{Z.L. Wang}
 \email{ziliang.wang@kit.edu}

 \affiliation{Institute for Theoretical Physics,
 Karlsruhe Institute of Technology (KIT),\\
 76128 Karlsruhe, Germany\\}
\vspace*{5mm}

\begin{abstract}
\vspace*{1mm}\noindent
We derive the equations of motion for scalar metric perturbations
in a particular nonsingular boun\-cing cosmology,
where the big bang singularity is replaced by a spacetime defect
with a degenerate metric. The adiabatic perturbation solution
is obtained for nonrelativistic hydrodynamic matter.
We get the same result by working with conformal coordinates.
This last method is also valid for vector and tensor metric perturbations,
and selected results are presented.
We, finally, discuss several new effects from the linear perturbations
of this nonsingular bouncing cosmology,
such as across-bounce information transfer
and the possible imprint on cosmological perturbations
from a new phase responsible for the effective spacetime defect.
\vspace*{10mm}
\end{abstract}

\pacs{04.20.Cv, 98.80.Bp, 98.80.Jk}
\keywords{general relativity, big bang theory,
          mathematical and relativistic aspects of cosmology}

\maketitle

\section{Introduction}
\label{sec:Introduction}

The expanding universe appears to be described reasonably well
by the Friedmann solution~\cite{Friedmann1922-1924,MisnerThorneWheeler2017}.
This solution has, however, a big bang singularity
with diverging curvature
and energy density. Recently, it has been
shown~\cite{Klinkhamer2019} that the Friedmann big bang
singularity can be replaced by a spacetime defect, where
the spacetime metric is degenerate
but the curvature finite, as is the energy density
(further details are given in Ref.~\cite{Klinkhamer2019-More}).

This ``regularized'' big bang singularity
corresponds, in fact, to a nonsingular
\mbox{bounce~\cite{KlinkhamerWang2019-PRD,KlinkhamerWang2019-LHEP}.}
There is then a prebounce
phase where the positive cosmic scale factor decreases,
the bounce moment at which the positive cosmic scale factor is stationary,
and a postbounce phase where the positive cosmic scale factor
increases again. This particular nonsingular bouncing cosmology
is obtained from an extended version of general relativity,
which allows for degenerate metrics
(further remarks and references will be given in
Sec.~\ref{sec:Background-metric-and-perturbations}).

The urgent question, now, is if the bounce is
stable under small perturbations of the metric and the matter.
Perturbations are, therefore, the main topic of this article.
The focus will be on scalar metric perturbations, while
vector and tensor metric perturbations
are briefly mentioned in one of the Appendices.  %%PROOF1
Having obtained the behavior of the metric perturbations,
it is also possible to address the issue of information
transfer across the bounce.

At this moment, there is an important
point that we would like to make.
The degenerate-metric \textit{Ansatz}~\cite{Klinkhamer2019}
gives modified Friedmann equations which correspond
to \mbox{\emph{singular}} differential equations.
Even though these modified Friedmann equations have a nonsingular
solution for the background spacetime, there is still
the potential danger of singularities appearing in
perturbations of the metric. In this article,
we will find that also the metric perturbations have
nonsingular solutions, which is a nontrivial result
(different and potentially catastrophic behavior
has been found in a dynamic-vacuum-energy
model without big bang curvature
singularity~\cite{KlinkhamerLing2019,KlinkhamerWang2019-Instability}).

The outline of our article is now as follows.
In Sec.~\ref{sec:Background-metric-and-perturbations},
we review the degenerate-metric \textit{Ansatz}
(together with the heuristics of the resulting bounce solution)
and discuss general metric perturbations.
In Sec.~\ref{sec:Scalar-metric-perturbations},
we derive the equations of motion for scalar metric perturbations
and get the adiabatic perturbation solution
for the case of nonrelativistic hydrodynamic matter.
In Secs.~\ref{sec:Bounce-stability}
and \ref{sec:Across-bounce-information-transfer},
we briefly discuss the issues of bounce stability
and across-bounce information transfer.
In Sec.~\ref{sec:Discussion}, we summarize our results and
discuss how they may be relevant to the generation of
a scale-invariant power spectrum of cosmological perturbations.
In Appendix~\ref{app:Details}, we provide certain details for
the calculation of Sec.~\ref{sec:Scalar-metric-perturbations}.
In Appendix~\ref{app:Perturbations-with-conformal-coordinates},
we rederive our results for scalar metric perturbations
by use of conformal coordinates, and also give some results
for vector and tensor metric perturbations.
In Appendix~\ref{app:Imprint-from-a-new-phase?},
we present a scenario of how a possible new phase
(``quantum spacetime'' or something entirely different)
may leave an imprint on the spectrum of
cosmological perturbations.

%%\newpage%%tmp
\section{Background metric and perturbations}
\label{sec:Background-metric-and-perturbations}

The modified spatially flat Robertson--Walker (RW)
metric is given by~\cite{Klinkhamer2019}%
\bsubeqs\label{eq:mod-RW}
\beqa\label{eq:mod-RW-ds2}
\hspace*{-6mm}
ds^{2}\,\Big|_\text{mod.\;RW}
&\equiv&
g_{\mu\nu}(x)\, dx^\mu\,dx^\nu \,\Big|_\text{mod.\;RW}
=
- \frac{ t^{2}}{b^{2}+ t^{2}}\,d t^{2}
+ a^{2}( t )
\;\delta_{ij}\,dx^{i}\,dx^j\,,
\\[2mm]
\hspace*{-6mm}
b^2 &>& 0\,,
\\[2mm]
\label{eq:cosmic-scale-factor}
\hspace*{-6mm}
a( t ) &\in& \mathbb{R}\,,
\\[2mm]
\label{eq:range-cosmic-time-coordinate}
\hspace*{-6mm}
 t  &\in& (-\infty,\,\infty)\,,
\\[2mm]
\hspace*{-6mm}
x^{i} &\in& (-\infty,\,\infty)\,,
\eeqa
\esubeqs
where we set $c=1$ and let the spatial indices $i $, $j $
run over $\{1,\, 2,\, 3 \}$.
The  cosmic time coordinate was denoted ``$T$'' in
Refs.~\cite{Klinkhamer2019,Klinkhamer2019-More}, but, here, we simply
write  ``$ t $,'' while emphasizing that the coordinate range
is given by \eqref{eq:range-cosmic-time-coordinate}.

The metric from \eqref{eq:mod-RW} is degenerate
(with a vanishing determinant at $ t =0$) and describes
a spacetime defect with characteristic length scale $b>0\,$;
see Ref.~\cite{Klinkhamer2019-JPCS} for a general review
of this type of spacetime defect and Ref.~\cite{KlinkhamerSorba2014}
for a detailed discussion of related mathematical aspects
(other mathematical aspects
of degenerate metrics have been
discussed in Ref.~\cite{Horowitz1991}).
The defect length scale $b$ in the metric \eqref{eq:mod-RW-ds2} will,
for the moment, be considered as an external parameter
(see Sec.~\ref{sec:Discussion} for further discussion).

At this moment, it may be helpful to recall some details of the
nonsingular bouncing cosmology obtained in
Refs.~\cite{Klinkhamer2019,Klinkhamer2019-More,KlinkhamerWang2019-PRD}.
Inserting the metric \textit{Ansatz} \eqref{eq:mod-RW-ds2}
into the Einstein gravitational field
equation~\cite{MisnerThorneWheeler2017} and taking
the energy-momentum tensor of a homogeneous perfect fluid
[with energy density $\rho(t)$, pressure $P(t)$, and
a constant equation-of-state parameter $w$],
the following modified Friedmann equations are
obtained:%
\bsubeqs\label{eq:mod-Friedmann-equations-abcd}
\beqa
\label{eq:mod-Friedmann-equation-a}
\hspace*{-5mm}&&
\left(1+ \frac{b^{2}}{t^{2}}\right)
\left( \frac{1}{a(t)}\,\frac{d a(t)}{d t} \right)^{2}
= \frac{8\pi G_N}{3}\;\rho(t)\,,
%\eeqa
\\[3mm]
%\beqa
\label{eq:mod-Friedmann-equation-b}
\hspace*{-5mm}&&
\left( 1+\frac{b^{2}}{t^{2}}\right)
\left[
\frac{1}{a(t)}\,\frac{d^2 a(t)}{d t^2}
+\frac{1}{2}\,\left( \frac{1}{a(t)}\,\frac{d a(t)}{d t} \right)^{2}
\right]
-\frac{b^{2}}{t^{3}}\,\frac{1}{a(t)}\,\frac{d a(t)}{d t}
=-4\pi G_{N}\,P(t)\,,
\eeqa
%\\[3mm]
\beqa
\label{eq:mod-Friedmann-equation-c}
\hspace*{-5mm}&&
\frac{d}{d a} \bigg[ a^{3}\,\rho(a)\bigg]+ 3\, a^{2}\,P(a)=0\,,
\\[3mm]
\label{eq:mod-Friedmann-equation-d}
\hspace*{-5mm}&&
\frac{P(t)}{\rho(t) } = w  = \text{const}\,,
\eeqa
\esubeqs
where $G_{N}$ is Newton's gravitational coupling constant and
where the last equation will later be specialized
the case of nonrelativistic matter with $w=0$.

The equations \eqref{eq:mod-Friedmann-equations-abcd}
have a bounce solution
which is perfectly smooth at $t=0$ as long as $b\ne 0$.
For the nonrelativistic-matter case ($w=0$),
this solution is given by
\bsubeqs\label{eq:regularized-Friedmann-asol-rhosol}
\beqa
\label{eq:regularized-Friedmann-asol}
a(t)\,\Big|_\text{bounce}^\text{(w=0)} &=&
\sqrt[3]{\frac{b^{2}+t^{2}}{b^{2}+t_{0}^{2}}}\,,
\\[2mm]
\label{eq:regularized-Friedmann-rhosol}
\rho(t)\,\Big|_\text{bounce}^\text{(w=0)} &=&
\rho_{0}\;\frac{b^{2}+t_{0}^{2}}{b^{2}+t^{2}}\,,
\eeqa
\esubeqs
with normalization $a(t_{0})=1$ at $t_{0}>0$
and boundary condition $\rho(t_{0}) = \rho_{0}>0$.
The bouncing behavior of the  positive scale factor $a(t)$
from \eqref{eq:regularized-Friedmann-asol} is manifest:
$a(t)$ decreases for negative $t$ approaching $t=0^{-}$,
the bounce occurs at $t=0$ with
a vanishing time derivative of $a(t)$ at $t=0$,
and $a(t)$ increases for positive $t$ moving away from $t=0^{+}$.
Similar nonsingular bounce solutions for other
equation-of-state parameters have been discussed in
Refs.~\cite{KlinkhamerWang2019-PRD,KlinkhamerWang2019-LHEP}
and the detailed dynamics of the bounce has been studied
analytically in  Ref.~\cite{Klinkhamer2019-More}.

The heuristics of this particular type of bounce solution
is as follows.
As noted in Sec.~II of Ref.~\cite{Klinkhamer2019-More},
the modified Friedmann
equations \eqref{eq:mod-Friedmann-equation-a}
and \eqref{eq:mod-Friedmann-equation-b} can
be rewritten as the standard Friedmann equations with
an additional effective energy density $\rho_\text{defect}$ and
an additional effective pressure $P_\text{defect}$.
Specifically, the resulting equations read
\bsubeqs\label{eq:mod-Friedmann-equations-rewritten}
\beqa
&&
\left( \frac{\dot{a}}{a}\right)^{2}
=\frac{8 \pi G_{N}}{3}\,\Big[\rho + \rho_\text{defect} \Big]\,,
\\[2mm]
&&
\left[
\frac{\ddot{a}}{a}
+ \frac{1}{2}\,\frac{\dot{a}^{2}}{a^{2}}\,
\right]
=-4\pi G_{N}\,\Big[P + P_\text{defect} \Big]\,,
\\[2mm]
&&
\rho_\text{defect}  \equiv
-\frac{b^2}{b^2+t^2}\,\rho\,,
\\[2mm]
&&
P_\text{defect} \equiv
-\frac{b^2}{b^2+t^2}\,
\left(P + \frac{1}{4\pi G_{N}}\frac{1}{t}\,\frac{\dot{a}}{a} \right)\,,
\eeqa
\esubeqs
where the overdot stands for the derivative with respect to $t$.
We observe that $\rho_\text{defect}$ and $P_\text{defect}$
have ``matter-induced'' terms
\big[proportional to the matter quantities $\rho$ and $P$, respectively\big] and that $P_\text{defect}$ also has a ``vacuum'' term
\big[proportional to $(4\pi G_{N})^{-1}\,t^{-1}\, \dot{a}/a$\big].
For the sum of the total effective energy density
$\rho_\text{total} \equiv \rho + \rho_\text{defect}$ and
the total effective pressure $P_\text{total} \equiv P+P_\text{defect}$,
we have%
\beq
\rho_\text{total} + P_\text{total}
= \frac{t^2}{b^2+t^2}\,\big(\rho + P\big)
- \frac{1}{4\pi G_{N}}\,\frac{b^2}{b^2+t^2}\,\frac{1}{t}\,\frac{\dot{a}}{a}\,.
\eeq
With finite values of $\rho$ and $P$ at the moment
of the bounce ($t=0$) and the near-bounce behavior
$a(t) \sim a_B + c_2 \,t^2$, for $a_B>0$ and $c_2>0$
(or $a_B<0$ and $c_2<0$), the following inequality holds:
\beq\label{eq:NEC-violation}
\Big[\rho_\text{total} + P_\text{total}\Big]_{t=0} < 0\,,
\eeq
which can be extended to a finite interval around $t=0$.
The inequality \eqref{eq:NEC-violation} corresponds
to an effective violation of the null energy condition
($\rho_\text{total} + P_\text{total} \geq 0$), in agreement
with the general discussion on bounces and energy conditions
in Ref.~\cite{NovelloBergliaffa2008}.

We now return to the general background metric
from \eqref{eq:mod-RW},
which  will be called the unperturbed metric.
Henceforth, a bar over a quantity denotes its unperturbed value.
The perturbed metric can then be written as
\beq\label{eq:def-perturbed-metric}
g_{\mu \nu}(x)\,\Big|^\text{(perturbed)}_\text{mod.\;RW}
=\overline{g}_{\mu \nu}( t )+h_{\mu \nu}(x)\,,
\eeq
where $h_{\mu \nu} = h_{\nu \mu}$ is a small perturbation
compared to the unperturbed metric $\overline{g}_{\mu \nu}$
from \eqref{eq:mod-RW-ds2}.
The spatially isotropic and homogeneous background allows us to
decompose the metric perturbations into scalars, divergenceless vectors,
and divergenceless traceless symmetric tensors~\cite{Mukhanov2005,Weinberg2008}.
The main focus of this article will be on scalar metric perturbations.

%%\newpage%%tmp
\section{Scalar metric perturbations}
\label{sec:Scalar-metric-perturbations}

\subsection{Metric Ansatz}
\label{subsec:Scalar-metric-perturbations}

The \textit{Ansatz} for the
metric with scalar perturbations is taken as follows:
\beqa\label{eq:metric-scalar-perturb.}
&&ds^{2}\,\Big|_\text{mod.\;RW}^\text{(scalar\;pert.)}=
\nonumber\\[1mm]
&&
-\big(1+E\big)\,\frac{ t^{2}}{b^{2}+ t^{2}}\,d t^{2}
+2\,\overline{a}\,\frac{\partial F}{\partial x^{i}}\,dx^{i}d t
+\overline{a}^{2}\left[\big(1+A\big)\,\delta _{ij}
+\frac{\partial B^{2}}{\partial x^{i} \partial x^{j}}\right]dx^{i}dx^{j}\,,
\eeqa
where the perturbations
$E,F,A$ and $B$ are functions of all spacetime coordinates
$\{ t ,\,x^1,\, x^{2},\, x^{3} \}$
and the background scale factor $\overline{a}$
is a function of only $ t $.

%%\newpage%%tmp
\subsection{Newtonian gauge}
\label{sec:Newtonian-gauge}

Consider the following transformation of the spacetime coordinates:
\beq\label{eq:gauge-trans.}
x^{\mu} \to \widetilde{x}^{\mu} =x^{\mu} +\xi^{\mu}\,,
\eeq
where the parameters $\xi^{\mu} \equiv \xi^{\mu} (x)$
are infinitesimal functions of the spacetime coordinates.
By decomposing the spatial part of $\xi^{\mu}$ into
the gradient of a spatial scalar
and a divergenceless vector~\cite{Mukhanov2005,Weinberg2008},
\bsubeqs
\beqa
\xi^{i} = \partial^{i}\, \xi_{S} + \xi_{V}^{i} \,,
\eeqa
\beqa
\partial_{i}\, \xi_{V}^{i}=0\,,
\eeqa
\esubeqs
we have the following transformations of the
metric functions from \eqref{eq:metric-scalar-perturb.}
under the change of coordinates \eqref{eq:gauge-trans.}:
\bsubeqs\label{scalar-gauge-trans.}
\beqa\label{scalar-gauge-trans.a}
\widetilde{E}=
E-\frac{2\,b^{2}}{ t }\,\xi^{0}
-\frac{\partial \xi^{0}}{\partial  t }\,,
\eeqa
\beqa\label{scalar-gauge-trans.b}
\widetilde{F}=
F-\frac{\overline{a}}{2}\,\frac{\partial \xi_{S}}{\partial  t }
+\frac{ t^{2}}{b^{2}+ t^{2}}\,\frac{\xi^{0}}{2\,\overline{a}}\,,
\eeqa
\beqa\label{scalar-gauge-trans.c}
\widetilde{A}=A-\frac{2\,\dot{\overline{a}}}{\overline{a}}\,\xi^{0}\,,
\eeqa
\beqa\label{scalar-gauge-trans.d}
\widetilde{B}=B-2\,\xi_{S}\,,
\eeqa
\esubeqs
where the overdot stands for the partial derivative with respect to $ t $.
Note that only $\xi^{0}$ and $\xi_{S}$ contribute to the transformations
of scalar metric perturbations.

Following Sec.~7.1.2 of Ref.~\cite{Mukhanov2005},
we  construct the following gauge-invariant quantities:
\bsubeqs
\beqa
2\,\Phi \equiv
E-\frac{\partial}{\partial  t }
\left[ 2\,\overline{a}\, \frac{b^{2}+ t^{2}}{ t^{2}}
\left( F-\frac{\overline{a}}{4}\,\dot{B}\right)\right]
-\frac{2\,b^{2}}{ t }
\left[ 2\,\overline{a}\,\frac{b^{2}+ t^{2}}{ t^{2}}
\left(F-\frac{\overline{a}}{4}\,\dot{B} \right)\right]\,,
\eeqa
\beqa
2\,\Psi \equiv A-4\,\dot{\overline{a}}\,\frac{b^{2}+ t^{2}}{ t^{2}}
\left(F-\frac{\overline{a}}{4}\,\dot{B}\right)\,.
\eeqa
\esubeqs
In this article, we will use the Newtonian gauge
(the origin of the name will become clear later on),
\beq\label{eq:def.newtonian gauge}
F=B=0,
\eeq
which can be reached by, first, choosing an appropriate $\xi_{S}$
in (\ref{scalar-gauge-trans.d}) and, then, an appropriate $\xi^{0}$
in (\ref{scalar-gauge-trans.b}).
In this gauge, the line element (\ref{eq:metric-scalar-perturb.})
reduces to
\beq\label{eq:newtonian-gauge-scalar-metric}
ds^{2}\,\Big|_\text{mod.\;RW}^{\text{(scalar\;pert.\;Newtonian-gauge)}}
=-\big(1+2\,\Phi\big)\,\frac{ t^{2}}{b^{2}+ t^{2}}\,d t^{2}
+\overline{a}^{2}\,\big(1+2\,\Psi\big)\,\delta _{ij}\, dx^{i}dx^{j}\,.
\eeq
Three remarks are in order.
First, note that, for later use,
the perturbations $\Phi$ and $\Psi$ are assumed to be spatially localized.
Second, remark that,
after choosing the Newtonian gauge,
there is no further freedom to make coordinate transformations,
while remaining within the
\textit{Ansatz} \eqref{eq:newtonian-gauge-scalar-metric}.
Third, observe that the perturbed metric
from \eqref{eq:newtonian-gauge-scalar-metric} still has,
for finite $\Phi$ and $\Psi$, a vanishing determinant at $ t =0$.

%%\newpage%%tmp
\subsection{Hydrodynamic matter perturbations}
\label{subsec:Hydrodynamic-matter-perturbations}

\subsubsection{General results}
\label{subsubsec:General-results}

Now, consider a perfect fluid with energy-momentum tensor
\beq\label{eq:EMT-perfect-fluid}
T_{\mu \nu}=P\,g_{\mu \nu} +\big(\rho +P\big)\,U_{\mu} U_{\nu}\,,
\eeq
where $P$ is the pressure, $\rho$ the energy density, and
$U^{\mu}$ the four-velocity. With the perturbed metric
(\ref{eq:newtonian-gauge-scalar-metric}),
the first-order perturbations of the $00$ and $ij$ components
of the energy-momentum tensor are
\bsubeqs\label{eq:perturb-EMT}
\beqa
\delta T _{00}=
2 \, \frac{ t^{2}}{b^{2}+ t^{2}}\,\overline{\rho}\,\Phi
+\frac{ t^{2}}{b^{2}+ t^{2}}\,\delta\rho\,,
\eeqa
\beqa
\delta T_{ij}=
2\,\overline{a}^{2}\, \overline{P}\, \Psi\,\delta _{ij}
+\overline{a}^{2}\,\delta P\,\delta _{ij}\,,
\eeqa
\esubeqs
which gives
\bsubeqs\label{eq:perturb-EMT-mix}
\beqa
\delta T^{0}_{\;\,0}=-\delta \rho\,,
\eeqa
\beqa
\delta T^{i} _{\;j}=\delta P\;\delta^{i}_{\;j}\,.
\eeqa
\esubeqs
%\bsubeqs\label{eq:perturb-EMT}  %%ZLW %%FRK: keep tmp
%\beqa
%\delta T _{00}=-\overline{\rho}h_{00}+\frac{ t^{2}}{b^{2}+ t^{2}}\delta \rho\,,
%\eeqa
%\beqa
%\delta T_{ij}=\overline{P}h_{ij}+\overline{a}^{2}\delta _{ij}\delta P\,,
%\eeqa
%\beqa
%\delta T_{0i} =\overline{P}h_{0i}-(\overline{\rho}+\overline{P})\sqrt{\frac{ t^{2}}{b^{2}+ t^{2}}}\delta U_{i}\,,
%\eeqa
%\esubeqs
%where the Latin indices $i $, $j $ run over the three spatial coordinates.

The unperturbed energy density $\overline{\rho}( t )$
and pressure $\overline{P}( t )$ are determined
by the following modified Friedmann
equations~\cite{Klinkhamer2019,Klinkhamer2019-More}:
\bsubeqs\label{eq:unperturbed modified friedmann equations}
\beq
\left(1+\frac{b^{2}}{ t^{2}} \right)
\left( \frac{\dot{\overline{a}}}{\overline{a}}\right)^{2}
=\frac{8 \pi G_{N}}{3}\,\overline{\rho}\,,
\eeq
\beq
\left( 1+\frac{b^{2}}{ t^{2}}\right)
\left[
\frac{2\,\ddot{\overline{a}}}{\overline{a}}
+ \frac{\dot{\overline{a}}^{2}}{\overline{a}^{2}}\,
\right]
-\frac{2\,b^{2}}{ t ^{3}}\,\frac{\dot{\overline{a}}}{\overline{a}}
=-8\pi G_{N}\,\overline{P}\,,
\eeq
\esubeqs
which were already given as \eqref{eq:mod-Friedmann-equation-a}
and \eqref{eq:mod-Friedmann-equation-b}
in Sec.~\ref{sec:Background-metric-and-perturbations}.
From a direct calculation of the perturbed Einstein tensor for
the perturbed metric (\ref{eq:newtonian-gauge-scalar-metric}),
together with (\ref{eq:perturb-EMT-mix}) and
(\ref{eq:unperturbed modified friedmann equations}),
we get the following equations of motion for scalar metric perturbations:
\bsubeqs\label{eq:EOM-for-scalar-perturbations}
\beqa\label{eq:density-perturbation-newtonian-gauge}
4\pi G_{N} \,\delta \rho
&=&
\frac{\triangle \Phi}{\overline{a}^{2}}
-3\,\frac{\dot{\overline{a}}^{2}}{\overline{a}^{2}}\,
\frac{b^{2}+ t^{2}}{ t^{2}}\,\Phi
-3\,\frac{\dot{\overline{a}}}{\overline{a}}\,\frac{b^{2}
+ t^{2}}{ t^{2}}\,\dot{\Phi}\,,
\\[2mm]
\label{eq:pressure-perturbation-newtonian-gauge}
4\pi G_{N} \,\delta P  &=&
\frac{b^{2}+ t^{2}}{ t^{2}}\,\ddot{\Phi}
+\frac{b^{2}+ t^{2}}{ t^{2}}
\left(\frac{\dot{\overline{a}}^{2}}{\overline{a}^{2}}
     +\frac{2\,\ddot{\overline{a}}}{\overline{a}}\right)\Phi
+4\,\frac{\dot{\overline{a}}}{\overline{a}}\,
\frac{b^{2}+ t^{2}}{ t^{2}}\,\dot{\Phi}
-2\,\frac{b^{2}}{ t ^{3}}\,\frac{\dot{\overline{a}}}{\overline{a}}\,
\Phi -\frac{b^{2}}{ t ^{3}}\,\dot{\Phi}\,,
\nonumber\\[1mm]&&
\eeqa
\esubeqs
where $\triangle$ is the Laplace operator in three-dimensional Euclidean space
and where we have used
\beq\label{eq:Psi-is-minus-Phi}
\Psi =-\Phi \,,
\eeq
which follows from the perturbed off-diagonal spatial Einstein equation.
Details of this calculation are relegated to Appendix~\ref{app:Details}.
Note that \eqref{eq:density-perturbation-newtonian-gauge}
for constant $\overline{a}( t )$ reproduces the Poisson equation
of Newtonian gravity, which explains the name of the
gauge~\cite{Mukhanov2005}.

Considering adiabatic perturbations, we have
\beq\label{eq:adiabatic perturbation}
\delta P= c^{2}_{s} \, \delta \rho \,,
\eeq
where $c^{2}_{s}$ is the square of the speed of sound~\cite{Mukhanov2005}.
Combining (\ref{eq:EOM-for-scalar-perturbations})
and (\ref{eq:adiabatic perturbation}),
we get the following equation of motion for $\Phi( t ,\,\mathbf{x})$:
\beqa\label{eq:eom-Phi}
&&
\frac{b^{2}+ t^{2}}{ t^{2}}\,\ddot{\Phi}
-c^{2} _{s}\,\frac{\triangle \Phi}{\overline{a}^{2}}
+\frac{b^{2}+ t^{2}}{ t^{2}}
\left[\frac{\dot{\overline{a}}^{2}}{\overline{a}^{2}}\,\Big(1+3\,c^{2}_{s}\Big)
+\frac{2\,\ddot{\overline{a}}}{\overline{a}}\right]\Phi
+\frac{\dot{\overline{a}}}{\overline{a}}\,
\frac{b^{2}+ t^{2}}{t^{2}}\,\Big(4+3\,c^{2}_{s}\Big)\,\dot{\Phi}
\nonumber\\[1mm]
&&-2\,\frac{b^{2}}{ t^{3}}\,
\frac{\dot{\overline{a}}}{\overline{a}}\,\Phi
-\frac{b^{2}}{t^{3}}\,\dot{\Phi} = 0\,,
\eeqa
which is the basic equation for adiabatic perturbations.

Equations (\ref{eq:EOM-for-scalar-perturbations}) and (\ref{eq:eom-Phi})
are \emph{singular} differential equations
(the singularity appears at $ t = 0$),
but they have \emph{nonsingular}
solutions that will be presented shortly.
As mentioned in Sec.~\ref{sec:Introduction},
the same behavior has been observed
for the differential equations and solutions of the background
spacetime~\cite{Klinkhamer2019,Klinkhamer2019-More}.

\subsubsection{Nonrelativistic matter}
\label{subsubsec:Nonrelativistic-matter}

For nonrelativistic matter, we have
the following background solution
from \eqref{eq:regularized-Friedmann-asol-rhosol}:%
\bsubeqs\label{eq:tips-matter-scalar pertur.}
\beqa
P&=&c^{2} _{s}=0\,,
\\[2mm]
\overline{\rho}( t )&\propto& \big[\overline{a}( t )\big]^{-3}\,,
\\[2mm]
\label{eq:sol.scale factor-matter case}
\overline{a}( t )&=&\sqrt[3]{\frac{b^{2}+ t^{2}}{b^{2}+ t _{0}^{2}}}\,,
\eeqa
\esubeqs
where $\overline{a}( t )$ has been normalized to unity
at $ t = t _{0}>0$\,.
In this case, (\ref{eq:eom-Phi}) has the solution
\beq
\label{eq:sol.Phi-matter-case}
\Phi( t ,\,\mathbf{x}) =
C_{1}(\mathbf {x})+
\frac{b^{5/3}\,C_{2}(\mathbf{x})}{\big(b^{2}+ t^{2}\big)^{5/6}}\,,
\eeq
where $C_{1} (\mathbf{x})$ and $C_{2} (\mathbf{x})$ are arbitrary
dimensionless functions of the spatial coordinates $\mathbf{x}$\,.
Notice that both modes in \eqref{eq:sol.Phi-matter-case}
are nonsingular at $ t =0$\,, which
will be discussed further in Sec.~\ref{sec:Bounce-stability}.

As a special case of \eqref{eq:sol.Phi-matter-case},
consider a plane-wave perturbation
with a single comoving wave vector $\mathbf{k}$,
\beq
\label{eq:C12-plane-wave}
C_{1,2}(\mathbf{x}) =\widehat{C}_{\mathbf{k},\,1,2}\;
\text{exp}\left(i\mathbf{k}\cdot\mathbf{x}\right)\,,
\eeq
where $\widehat{C}_{\mathbf{k},\,1}$
and $\widehat{C}_{\mathbf{k},\,2}$
are the dimensionless amplitudes. The amplitude of such
a plane-wave scalar metric perturbation is given by
\beq
\label{eq:amplitude-sol.Phi-matter-case}
\Phi_\mathbf{k}( t ) =
\widehat{C}_{\mathbf{k},\,1}
+\frac{b^{5/3}\,\widehat{C}_{\mathbf{k},\,2}}
      {\big(b^{2}+ t^{2}\big)^{5/6}}\,.
\eeq
From \eqref{eq:density-perturbation-newtonian-gauge},
the corresponding energy density perturbation has the
following amplitude:
\beqa\label{eq:sol.density-perturb.}
\frac{\delta \rho _\mathbf{k}( t )}{\overline{\rho}( t )}
&=&
-\left[ 2+\frac{3}{2}\, k^{2}\,
\big(b^{2}+ t _{0}^{2}\big)^{2/3}\,\big(b^{2}+ t^{2}\big)^{1/3}
\right]\widehat{C}_{\mathbf{k},\,1}
\nonumber\\[1mm]
&&
+\left[ 3-\frac{3}{2}\,k^{2}\,
\big(b^{2}+ t _{0}^{2}\big)^{2/3}\,\big(b^{2}
+ t^{2}\big)^{1/3} \right]
\frac{b^{5/3}\,\widehat{C}_{\mathbf{k},\,2}}{\big(b^{2}+ t^{2}\big)^{5/6}}\,,
\eeqa
with $k\equiv |\mathbf{k}|$. The perturbation results for different wave vectors
$\mathbf{k}$ can be superposed, in order to obtain localized wave packets.

For $t \ne 0$ and a physical wavelength much larger than the Hubble horizon
($1/H \equiv \overline{a}/\dot{\overline{a}}$),
\beq
\frac{\overline{a}^{2}}{k^{2}} \gg \frac{1}{H^{2}} >
\frac{ t^{2}/(b^{2}+ t^{2})}{H^{2}}\,,
\eeq
we have from \eqref{eq:sol.density-perturb.}
\bsubeqs\label{eq:long-short-wavelength-sol.density-perturb.}
\beq\label{eq:long-wavelength-sol.density-perturb.}
\left.\frac{\delta \rho_\mathbf{k}( t )}{\overline{\rho}( t )}
\right|^\text{(long-wavelength)}
\sim
-2\,\widehat{C}_{\mathbf{k},\,1}
+\frac{3\,\widehat{C}_{\mathbf{k},\,2}}{\big(1+ t^{2}/b^{2}\big)^{5/6}}\,.
\eeq
For a short physical wavelength, we get
\beq\label{eq:short-wavelength-sol.density-perturb.}
\left.\frac{\delta \rho_\mathbf{k}( t )}{\overline{\rho}( t )}
\right|^\text{(short-wavelength)}
\sim
-\frac{3}{2}\,k^{2}\,( t _{0}+b^{2})^{2/3}
\left[
\widehat{C}_{\mathbf{k},\,1}\,\sqrt[3]{b^{2}+ t^{2}}
+\frac{b^{2/3}\,\widehat{C}_{\mathbf{k},\,2}}{\sqrt{1+ t^{2}/b^{2}}}
\right]\,.
\eeq
\esubeqs
Remark that the growing mode in (\ref{eq:short-wavelength-sol.density-perturb.})
is proportional to the cosmic scale factor $\overline{a}( t )$
from (\ref{eq:sol.scale factor-matter case}),
just as happens for the standard matter-dominated Friedmann universe;
cf. Eq.~(7.56) of Ref.~\cite{Mukhanov2005}.

%The results \eqref{eq:long-wavelength-sol.density-perturb.}
%and \eqref{eq:short-wavelength-sol.density-perturb.}
%will be re-derived with
The results \eqref{eq:long-wavelength-sol.density-perturb.}
and \eqref{eq:short-wavelength-sol.density-perturb.}
will be rederived with an auxiliary cosmic time coordinate
in Appendix~\ref{app:Perturbations-with-conformal-coordinates},
which also contains results on vector and tensor metric perturbations.

%%\newpage%%tmp
\section{Bounce stability}
\label{sec:Bounce-stability}

The results from Sec.~\ref{subsubsec:Nonrelativistic-matter}
show that plane-wave scalar
metric perturbations and the corresponding
adiabatic density perturbations are \emph{finite}
at the moment of the bounce, $ t =0$.
Specifically, these results are given by
\eqref{eq:amplitude-sol.Phi-matter-case}
and \eqref{eq:sol.density-perturb.},
for an arbitrary comoving wave vector $\mathbf{k}$.

But the \emph{magnitude} of the metric perturbations
at $ t =0$ must also be small enough,
so as to keep the background metric essentially unchanged.
For scalar metric perturbations in the Newtonian gauge,
the perturbed metric is given by \eqref{eq:newtonian-gauge-scalar-metric}.
In order to keep the bounce essentially unchanged,
the absolute value of both perturbations
in \eqref{eq:newtonian-gauge-scalar-metric} must be
significantly less than unity at $ t =0$,%
\bsubeqs\label{eq:C1-C2-bounds}
\beqa
\big|2\,\Phi(0,\,\mathbf{x})\big| &\ll&  1\,,
\\[2mm]
\big|2\,\Psi(0,\,\mathbf{x})\big| &\ll&  1\,.
\eeqa
\esubeqs
With the solutions \eqref{eq:Psi-is-minus-Phi}
and \eqref{eq:amplitude-sol.Phi-matter-case},
there are then the following bounds on the amplitudes of the
plane-wave  scalar metric perturbations:
\bsubeqs\label{eq:C1hat-C2hat-bounds}
\beqa
\big|2\,\widehat{C}_{\mathbf{k},\,1}\big|  &\ll&  1\,,
\\[2mm]
\big|2\,\widehat{C}_{\mathbf{k},\,2}\big| &\ll&  1\,.
\eeqa
\esubeqs
In other words, having small enough amplitudes $|\widehat{C}_{\mathbf{k},\,1}|$
and $|\widehat{C}_{\mathbf{k},\,2}|$ does not disturb the bounce.
As mentioned in Sec.~\ref{sec:Introduction},
a similar conclusion does not hold for a
dynamic-vacuum-energy model~\cite{KlinkhamerLing2019}, which is
found to have a violent instability~\cite{KlinkhamerWang2019-Instability}
at the moment of the big bang, where the cosmic scale factor
vanishes.

Returning to our degenerate-metric bounce,
there is, however, a puzzle.
Consider, at $ t = t _\text{start}<0$  in the prebounce phase,
the hypothetical generation of plane-wave scalar metric perturbations
with an initial amplitude $\widehat{\Phi}_{\mathbf{k},\,\text{start}}$.
Generically, both modes
in \eqref{eq:amplitude-sol.Phi-matter-case} are exited,
but let us focus on the nonconstant mode, so that we have
\beq\label{eq:Phi-sol-with-bc-at-tstart}
\Phi_\mathbf{k}( t )
\sim
\widehat{\Phi}_{\mathbf{k},\,\text{start}}\,
\frac{(b^2+ t _\text{start}^2)^{5/6}}{(b^2+ t ^2)^{5/6}}\,.
\eeq
Perturbations generated at $ t _\text{start} \ll -b <0$
will grow with cosmic time $ t $ and give at the moment
of the bounce,  $ t =0$,
\beq\label{eq:bounce-value-of-Phi-sol-with-bc-at-tstart}
\Phi_\mathbf{k}(0)
\sim
\widehat{\Phi}_{\mathbf{k},\,\text{start}}\,
\frac{(b^2+ t _\text{start}^2)^{5/6}}{b^{5/3}}
\sim
\widehat{\Phi}_{\mathbf{k},\,\text{start}}\,
\frac{| t _\text{start}|^{5/3}}{b^{5/3}}\,.
\eeq
The stability condition \eqref{eq:C1-C2-bounds}
then requires an extremely small start amplitude,
\beq\label{eq:Phi-start-bound}
|\widehat{\Phi}_{\mathbf{k},\,\text{start}}| \ll
\frac{b^{5/3}}{| t _\text{start}|^{5/3}}\,.
\eeq
The puzzle, now, is how to guarantee a sufficiently
small amplitude $|\widehat{\Phi}_{\mathbf{k},\,\text{start}}|$
if the scalar metric perturbations are generated in the prebounce phase at
early times, $ t _\text{start} \ll -b <0$.

%%\newpage%%tmp
The puzzle outlined in the previous paragraph is, of course, well known
to practitioners of nonsingular-bouncing-cosmology scenarios
(cf. Refs.~\cite{Buchbinder-etal2007,Quintin-etal2014,%
LevyIjjasSteinhardt2015,BrandenbergerPeter2016,IjjasSteinhardt2018}
and references therein).
In fact, this is the motivation for postulating an exotic component
in the prebounce phase, with an equation-of-state parameter
$w_\text{ex} \equiv P_\text{ex}/\rho_\text{ex} \geq 1$.
Approaching the bounce from the prebounce side,
the energy density of an exotic component with $w_\text{ex}>1$
``grows to dominate all other forms of
energy, including inhomogeneities, anisotropy and spatial
curvature'' (quote from Ref.~\cite{LevyIjjasSteinhardt2015}).
For this reason, we have also
considered~\cite{KlinkhamerWang2019-PRD,KlinkhamerWang2019-LHEP}
asymmetric versions of the degenerate-metric bounce
with $w_\text{prebounce}=1$ and $w_\text{postbounce}\sim 1/3$.

An alternative physical interpretation of the
degenerate-metric bounce is as follows.
The thermodynamic arrow of time may be considered to run
in the direction of growing density perturbations
(cf. the discussion in Appendix~B of Ref.~\cite{KlinkhamerLing2019},
which contains further references).
From our result \eqref{eq:short-wavelength-sol.density-perturb.}
on the short-wavelength growing mode,
we can then define the thermodynamic
time $\mathcal{T}( t )$ as
$\mathcal{T}( t )\equiv t $ for $ t  > 0$ and
$\mathcal{T}( t )\equiv- t $ for $ t  \leq 0$.
In short, we have $\mathcal{T}( t ) \equiv | t |$.
Now assume that small metric perturbations
can \emph{only} be generated near the bounce at $\mathcal{T}=0$.
Then, the corresponding matter perturbations will simply grow
with $\mathcal{T} \equiv| t |$.
In other words, they grow equally on both ``sides'' of the bounce,
not endangering the bounce any further.
The physical motivation of this alternative scenario
will be discussed in the last paragraph of Sec.~\ref{sec:Discussion}.

%%\newpage%%tmp
\section{Across-bounce information transfer}
\label{sec:Across-bounce-information-transfer}

Having established the propagation of scalar metric
perturbations in Sec.~\ref{subsubsec:Nonrelativistic-matter}
and of tensor metric perturbations
in Appendix~\ref{app:Vector-tensor-metric-perturbations},
we are able to address the question of information transfer
\emph{across} the bounce.

Consider a plane-wave scalar metric perturbation triggered at
\beq
\label{eq:t-start}
 t = t _\text{start}<0\,,
\eeq
with amplitude
\beq
\label{eq:amplitude-Phi-at-tstart}
\Phi_\mathbf{k}( t _\text{start}) =
\widehat{\Phi}_{\mathbf{k},\,\text{start}}
\eeq
and an appropriate nonvanishing value of
$\dot{\Phi}_\mathbf{k}( t _\text{start})$
\big[from \eqref{eq:Phi-sol-with-bc-at-tstart},
we have
$\dot{\Phi}_\mathbf{k}( t _\text{start})/\widehat{\Phi}_{\mathbf{k},\,\text{start}}
\sim  -(5/3)\, t _\text{start}/(b^2+ t _\text{start}^2)$\big].
According to the solution \eqref{eq:Phi-sol-with-bc-at-tstart},
this perturbation grows until the bounce at $ t =0$
is reached and then decreases as $ t $ increases further.

Taking the observation time symmetrically for illustrative purposes,
\beq
\label{eq:t-obs}
 t _\text{obs} = - t _\text{start}>0\,,
\eeq
we find, from \eqref{eq:Phi-sol-with-bc-at-tstart},
the following observed perturbation amplitude:
\beq
\label{eq:amplitude-Phi-at-tobs}
\Phi_\mathbf{k}( t _\text{obs}) =
\widehat{\Phi}_{\mathbf{k},\,\text{start}}\,.
\eeq
In this way, we are, in principle, able to transfer
information across the bounce.
With a sequence of scalar-metric-perturbation pulses, for example,
it is possible to compose a message in Morse code that
starts in the prebounce phase, passes across the bounce,
and ends up in the postbounce phase.

More realistic would be to use short-wavelength gravitational waves,
and the same result on across-bounce information transfer
is obtained from
\eqref{eq:tensor-metric-perturbations-hij-short-wavelengths}.
In fact, we have already discussed in Ref.~\cite{KlinkhamerWang2019-PRD}
the across-bounce effects
of ``gravitational standard candles,''
so that the results of this article fill in one of the missing details
(the other missing detail [\textit{sic}]
is the actual existence of these standard candles).

%%\newpage%%tmp
\section{Discussion}
\label{sec:Discussion}

In the present article, we have obtained first results for
the metric
perturbations of a particular nonsingular bouncing cosmology.
This type of bounce~\cite{Klinkhamer2019}
relies on an extension of
general relativity, namely by allowing for degenerate
metrics (see, e.g., Ref.~\cite{Horowitz1991}
for an earlier discussion of general relativity
with degenerate metrics).
In fact, our degenerate-metric \emph{Ansatz}
describes what may be called a ``spacetime defect''
(see Ref.~\cite{Klinkhamer2019-JPCS} for a review).
The matter content of the corresponding nonsingular bouncing cosmology
is entirely standard, without any problem whatsoever
as regards unitarity and microcausality
(see, e.g., Sec.~III of Ref.~\cite{BrandenbergerPeter2016}
for an overview of other bounce realizations, some with potentially
problematic matter content).
In our case, singularity theorems are  circumvented by having
a degenerate metric with a vanishing
determinant on a three-dimensional submanifold (further discussion and
references can be found in Ref.~\cite{Klinkhamer2019}).

The main open question for our degenerate-metric bounce
is the physical origin of the corresponding spacetime
defect, notably its length scale $b$
in the metric  \emph{Ansatz} \eqref{eq:mod-RW}.
It could very well be that the spacetime defect at $t=0$
(in the notation of our \emph{Ansatz})
traces back to a new phase.
But nothing is known for sure about such a phase.
For the moment, we make no assumption about the
physical origin of the defect length scale $b$ and only require
that the numerical value of $b$ is large enough, so that
Einstein's classical gravity holds.

As it stands, we have with the degenerate-metric
\emph{Ansatz} \eqref{eq:mod-RW} from Ref.~\cite{Klinkhamer2019}
an economic way to describe a nonsingular bounce,
assuming such a bounce to be relevant to our Universe.
In that case, it is worthwhile to study the perturbations of the metric,
and we have initiated that analysis in the present article.

For any cosmological model
aiming to describe the evolution of the very early universe,
it is crucial to be able to produce a scale-invariant power spectrum
of cosmological perturbations.
The post-big-bang exponential expansion
of the inflationary scenario
does the job~\cite{Mukhanov2005,Weinberg2008}.
But prebounce or ekpyrotic scenarios
can also get a scale-invariant power spectrum
of fluctuations, using either axions or an additional scalar field
(see, e.g., Sec.~II of Ref.~\cite{BrandenbergerPeter2016}).
Recently, there have been effective-field-theory calculations
of  linear perturbations in
null-energy-condition-violating bounce models;
see Refs.~\cite{Creminelli2016,Cai-etal2016a,Cai-etal2016b}
and references therein.

Let us return to the degenerate-metric bounce~\cite{Klinkhamer2019},
where the matter obeys the standard energy conditions.
Even though the matter-dominated prebounce contraction phase
considered in this article is far from perfect
(as discussed in Sec.~\ref{sec:Bounce-stability}),
vacuum fluctuations in the prebounce phase appear to
be converted into a scale-invariant power spectrum
of density perturbations
in the postbounce phase
(cf. Sec.~II B of Ref.~\cite{BrandenbergerPeter2016}).
The actual mechanism which generates the required
scale-invariant power spectrum of cosmological perturbations may
depend on certain details of the prebounce phase
(cf. Refs.~\cite{LevyIjjasSteinhardt2015,IjjasSteinhardt2018}).
But, regardless of the prebounce generation mechanism
of the cosmological perturbations,
our explicit degenerate-metric model of the bounce,
with the time-symmetric results \eqref{eq:amplitude-sol.Phi-matter-case},
\eqref{eq:sol.density-perturb.},
and \eqref{eq:tensor-metric-perturbations-hij-short-wavelengths},
has shown that these particular perturbations
are unaffected by the dynamics of the bounce itself.
In our degenerate-metric bounce scenario (with a large enough
value of the defect length scale $b$,
so as to remain in the classical-gravity phase),
the cosmological perturbations
from the prebounce phase can safely cross the bounce,
and the present article has shown this by concrete examples.

If, however, the defect length scale $b$ of our metric \eqref{eq:mod-RW}
is relatively small and related to an entirely new phase
(``quantum spacetime'' or something different),
then another scenario may be envisioned, which was already mentioned
in the last paragraph of Sec.~\ref{sec:Bounce-stability}.
Taking nonrelativistic matter as a toy model,
we obtain from the density perturbation
result \eqref{eq:sol.density-perturb.} at $ t =0$
that the perturbations have a critical comoving wavelength
$\lambda_\text{crit} \sim (c\,t_0)^{2/3}\,(b)^{1/3}$
(the corresponding physical wavelength is
$\widetilde{\lambda}_\text{crit} \sim b$).
If the perturbations (all or only part of them)
are generated by the new phase
and emerge at the classical time $ t =0$, then we may
expect to observe a different behavior for wavelengths below or above
this critical wavelength
(see Appendix~\ref{app:Imprint-from-a-new-phase?}
for a possible scenario).
This would, in principle, provide a way to determine the numerical
value of $b$~\cite{Endnote:previous-referee-2019},
in addition to the \textit{Gedankenexperiment}
presented in Ref.~\cite{KlinkhamerWang2019-PRD}.
As mentioned before, the crucial open question is
the physical origin of the defect length scale $b$,
related to a possible new phase or not.
But, even if the defect length scale $b$ is related to
such a new phase, this does not necessarily imply
that $b$ is of the order of the Planck length~\cite{Klinkhamer2007}.

%%%%\newpage%%tmp
\vspace*{-0mm}
\begin{acknowledgments}
\vspace*{-0mm}
The work of Z.L.W. is supported by the China Scholarship Council.
\end{acknowledgments}

%%\newpage%%tmp
\begin{appendix}

\section{Calculational details}
\label{app:Details}

In this Appendix,  we provide some details for the calculation of
\eqref{eq:EOM-for-scalar-perturbations} and \eqref{eq:Psi-is-minus-Phi} in
Sec.~\ref{subsubsec:General-results}.

From \eqref{eq:newtonian-gauge-scalar-metric} and
\eqref{eq:perturb-EMT-mix}, we get the following perturbed Einstein
equations (up to first-order perturbations):
\bsubeqs\label{eq:appendix-perturbed-einstein-equation}
\beqa\label{eq:appendix-00-perturbed-einstein-equation}
\hspace*{-6mm}
8\pi G_{N}\; \Big(\overline{\rho}+2\,\Phi\,\overline{\rho} +\delta \rho\Big) 
&=&
3\,\frac{b^2+t^2}{t^2}\, \frac{\dot{\overline{a}}^2}{\overline{a}^2}
+ 6\, \frac{b^2+t^2}{t^2}\,\frac{\dot{\overline{a}}}{\overline{a}}\,\dot{\Psi}
- \frac{2\,\triangle \Psi}{\overline{a}^2}\,,
\eeqa
%%\\[2mm]
\beqa
\label{eq:appendix-ij-perturbed-einstein-equation}
\hspace*{-6mm}
8\pi G_{N}\;\Big(\bar{P}+2\,\Psi\,\bar{P} +\delta{P}\Big) \, \delta _{ij}
&=&
\Bigg[\frac{2\,b^2}{t^{3}}\,\dot{\Psi}-2\,\frac{b^2+t^2}{t^2}\,\ddot{\Psi}
-6\,\frac{b^2+t^2}{t^2}\,\frac{\dot{\overline{a}}}{\overline{a}}\,\dot{\Psi} 
+ 2\,\frac{b^2+t^2}{t^2}\,\frac{\dot{\overline{a}}}{\overline{a}}\,\dot{\Phi} 
\nonumber\\[2mm]
\hspace*{-6mm}&&
+2\,\left(\frac{b^2}{t^{3}}\,\frac{\dot{\overline{a}}}{\overline{a}}
-\frac{b^2+t^2}{t^2}\,\frac{\ddot{\overline{a}}}{\overline{a}} \right)\,
\big(1+2\,\Psi -2\,\Phi\big)
+\frac{\triangle \big(\Phi + \Psi\big)}{\overline{a}^2}
\nonumber\\[2mm]
\hspace*{-6mm}&&
- \frac{b^2 +t^2}{t^2}\,\frac{\dot{\overline{a}}^2}{\overline{a}^2}\,
\big(1+2\,\Psi -2\,\Phi\big) \Bigg]\,\delta _{ij}
-\frac{1}{{\overline{a}^2}}\,\frac{\partial ^2}{\partial x^i \partial x^j}\,
\big(\Phi + \Psi\big)\,,
\nonumber\\[2mm]
\hspace*{-6mm}&&
\eeqa
\esubeqs
where $\triangle$ is the Laplace operator in three-dimensional Euclidean space.

With the assumption that $\Phi$ and $\Psi$  vanish at spatial infinity,
we obtain from \eqref{eq:appendix-ij-perturbed-einstein-equation} for $i \neq j$
that
\beq\label{eq:appendix-Psi-is-minus-Phi}
\Psi = -\Phi \,,
\eeq
as stated in \eqref{eq:Psi-is-minus-Phi} of the main text.

Without perturbations, the leading order terms in
\eqref{eq:appendix-perturbed-einstein-equation} are precisely
the background
equations \eqref{eq:unperturbed modified friedmann equations}.
From the first-order perturbations in
\eqref{eq:appendix-perturbed-einstein-equation},
together with the background equations
\eqref{eq:unperturbed modified friedmann equations} and
the result \eqref{eq:appendix-Psi-is-minus-Phi}, we get
\eqref{eq:EOM-for-scalar-perturbations} of the main text.​

%%\newpage%%tmp
\section{Perturbations with conformal coordinates}
\label{app:Perturbations-with-conformal-coordinates}

\subsection{Scalar metric perturbations}
\label{app:Scalar-metric-perturbations}

The modified spatially flat
Robertson--Walker metric (\ref{eq:mod-RW-ds2}) can be written as
\bsubeqs
\beqa
\label{eq:appendix-ds^{2}-conformal}
ds^{2}\,\Big|_\text{mod.\;RW}
&=&
\Omega^{2}(\eta)\;\Big(-d \eta^{2}+\delta _{ij}\,dx^{i} dx^j\Big)\,,
\eeqa
where $\eta$ is the conformal time defined by
\beqa
\label{eq:def.conformal time}
\Omega(\eta)\, d \eta
&=& \sqrt{\frac{ t^{2}}{b^{2}+ t^{2}}}\; d t \,,
\eeqa
\esubeqs
for $ t \in \mathbb{R}$ and defect length scale $b>0$
(see below for further details on $\eta$).
Perturbations of the conformally flat metric
metric (\ref{eq:appendix-ds^{2}-conformal})
have been widely studied in the literature; see, in
particular,  Ref.~\cite{Mukhanov2005}.

For our nonsingular degenerate-metric bouncing cosmology,
the metric perturbation solutions
take the same form as in the standard hot-big-bang model
but now with $\eta$ given by (\ref{eq:def.conformal time}).
For example, the plane-wave adiabatic
scalar perturbations for nonrelativistic
hydrodynamical matter have the following solutions
given by Eqs.~(7.55) and (7.56) in Ref.~\cite{Mukhanov2005}:%
\bsubeqs\label{eq:sol.matter-perturbation-conformal-time}
\beqa
\label{eq:sol.matter-perturbation-conformal-time-long-wavelengths}
\left.\frac{\delta {\rho}_\mathbf{k}(\eta)}{\overline{\rho}(\eta)}
\right|^\text{(long-wavelength)}
&\sim&
-2\, \widehat{C}_{\mathbf{k},\,1}
+ 3\, \widehat{C}_{\mathbf{k},\,2}\,\text{sgn}(\eta)\,\eta^{-5} \,,
\\[2mm]
\label{eq:sol.matter-perturbation-conformal-time-short-wavelengths}
\left.\frac{\delta {\rho}_\mathbf{k}(\eta)}{\overline{\rho}(\eta)}
\right|^\text{(short-wavelength)}
&\sim&
 -\frac{k^{2}}{6}\left(\widehat{C}_{\mathbf{k},\,1}\, \eta^{2}
 +\widehat{C}_{\mathbf{k},\,2}\,\text{sgn}(\eta)\,\eta^{-3} \right)\,,
\eeqa
\esubeqs
with $k\equiv |\mathbf{k}|$ and
constants $\widehat{C}_{\mathbf{k},\,1,2}$\,
(the extra sign factor multiplying $\widehat{C}_{\mathbf{k},\,2}$
is needed to get the correct boundary conditions
at the spacetime defect, as will be explained below).
For nonrelativistic matter, \eqref{eq:sol.scale factor-matter case}
and (\ref{eq:def.conformal time}) give
\bsubeqs\label{eq:Omega-eta}
\beqa
\label{eq:Omega}
\Omega(\eta) &=&  \frac{1}{9}\,\frac{\eta^{2}}{b^{2}+ t _{0}^{2}}  \,,
\\[2mm]
\label{eq:eta-matter-dominated-case}
\eta &=&
\begin{cases}
+ 3\,\sqrt[3]{b^{2}+ t _{0}^{2}}\;\;
  \sqrt[6]{b^{2}+ t^{2}\phantom{ t _{0}^{2}}}\,,
&  \;\;\text{for}\;\;  t  \geq 0\,,
 \\[1mm]
- 3\,\sqrt[3]{b^{2}+ t _{0}^{2}}\;\;
  \sqrt[6]{b^{2}+ t^{2}\phantom{ t _{0}^{2}}}\,,
&  \;\;\text{for}\;\;  t  \leq 0\,,
\end{cases}
\\[2mm]
\eta &\in& (-\infty,\,\eta_{-}]  \, \cup \, [\eta_{+},\,\infty)\,,
\\[2mm]
\eta_{\pm} &\equiv&
\pm \,3\;\sqrt[3]{b\,\big(b^{2}+ t _{0}^{2}\big)}\,,
\eeqa
\esubeqs
where the points $\eta=\eta_{-}$ and $\eta=\eta_{+}$ are identified
(in this way, the topology becomes $\mathbb{R}$).

The coordinate transformation \eqref{eq:eta-matter-dominated-case}
is not a diffeomorphism, so that the differential structure from
\eqref{eq:mod-RW-ds2} differs from that of
\eqref{eq:appendix-ds^{2}-conformal}.
In fact, $t$ from \eqref{eq:mod-RW} \emph{is} a good coordinate,
but $\eta$ from \eqref{eq:eta-matter-dominated-case} is \emph{not}
(there are different values $\eta_{\pm}$ for the single point $ t =0$).
Still, $\eta$ appears
to be a useful auxiliary coordinate away from the
spacetime defect at $\eta=\eta_{\pm}$. The
implication is that the $\eta$ domains $ (-\infty,\,\eta_{-})$
and $(\eta_{+},\,\infty)$ are disconnected, so that the boundary
conditions at $\eta =\eta_{\pm}$ require special care. See
Refs.~\cite{Klinkhamer2019,Klinkhamer2019-More,%
KlinkhamerSorba2014} for an extensive discussion of these issues.

At this moment, we remark that the extra minus signs for
the $\widehat{C}_{\mathbf{k},\,2}$ terms in
\eqref{eq:sol.matter-perturbation-conformal-time}
make for proper boundary conditions at $\eta =\eta_{\pm}$.
Indeed, inserting the $\eta$ expression from \eqref{eq:eta-matter-dominated-case}
into the perturbations
\eqref{eq:sol.matter-perturbation-conformal-time-long-wavelengths}
and \eqref{eq:sol.matter-perturbation-conformal-time-short-wavelengths},
we observe the resulting expressions to be even with respect to $ t $.
Introducing dimensionless constants $\widehat{c}_{\mathbf{k},\,1,\,2}$
by the definitions $\{\widehat{C}_{\mathbf{k},\,1},\,\widehat{C}_{\mathbf{k},\,2}\}
\equiv \{\widehat{c}_{\mathbf{k},\,1},\,b^5\,\widehat{c}_{\mathbf{k},\,2}\}$,
the final expressions are
\bsubeqs\label{eq:sol.matter-perturbation-from-eta-sol}
 \beqa
\hspace*{-0mm}
\left.\frac{\delta {\rho} _\mathbf{k}( t )}{\overline{\rho}( t )}
\right|^\text{(long-wavelength)}
&\sim&
 -2\, \widehat{c}_{\mathbf{k},\,1} + 3^{-4}\,b^5\,\widehat{c}_{\mathbf{k},\,2}\,
  \big(b^{2}+ t _{0}^{2}\big)^{-5/3} \,\big(b^{2}+ t^{2}\big)^{-5/6} \,,
\\[2mm]
\hspace*{-0mm}
\left.\frac{\delta {\rho} _\mathbf{k}( t )}{\overline{\rho}( t )}
\right|^\text{(short-wavelength)}
&\sim&
 -\frac{3}{2}\,k^{2}\big(b^{2}+ t _{0}^{2}\big)^{2/3}
 \Big[\, \widehat{c}_{\mathbf{k},\,1}
 \sqrt[3]{b^{2}+ t^{2}}
\nonumber\\[1mm]
\hspace*{-0mm}
&&
  +3^{-5}\,b^5\,\widehat{c}_{\mathbf{k},\,2}\,
 \big(b^{2}+ t _{0}^{2}\big)^{-5/3}\,\big(b^{2}
 + t^{2}\big)^{-1/2}\,\Big]\,,
\eeqa
\esubeqs
in agreement with our previous results
(\ref{eq:long-wavelength-sol.density-perturb.})
and (\ref{eq:short-wavelength-sol.density-perturb.}).

%%\newpage%%tmp
\subsection{Vector and tensor metric perturbations}
\label{app:Vector-tensor-metric-perturbations}

Results on vector and tensor metric perturbations
can be directly taken over
from Sec.~7.3.2 in Ref.~\cite{Mukhanov2005}, where,
for the nonrelativistic-matter case,
the conformal factor $a(\eta)$ is replaced by our
factor $\Omega(\eta)$ from \eqref{eq:Omega}
and $\eta$ is given by \eqref{eq:eta-matter-dominated-case}.

Here, we give some explicit results for the radiation-dominated case.
With the energy-momentum-tensor perturbation
$\delta T_i^0 = \Omega^{-1}\,(\overline{\rho} + \overline{P})\,\delta u_{\bot\, i}$
and the definition
$\delta v^{i} \equiv \Omega^{-1}\,\delta u_{\bot\, i}$,
the result from Eq~(7.94) in Ref.~\cite{Mukhanov2005} is
that plane-wave vector metric perturbations
for the radiation-dominated case
are constant with respect to the conformal time $\eta$,
\beq\label{eq:vector-metric-perturbations}
\delta v^{i}_\mathbf{k}
= \frac{\widehat{C}_{\mathbf{k},\,3}^{\;i}}{\Omega^{4}
\,\big(\overline{\rho} + \overline{P}\big)}
= \widehat{c}_{\mathbf{k},\,3}^{\;i}\,,
\eeq
where the last equality uses the
radiative behavior $\overline{P} =\overline{\rho}/3 \propto \Omega^{-4}$
and where the $\widehat{c}_{\mathbf{k},\,3}^{\;i}$
are appropriate dimensionless constants.

Turning to plane-wave  tensor metric perturbations
for the radiation-dominated case,
the result from Eq.~(7.98) in Ref.~\cite{Mukhanov2005} is
as follows:
\bsubeqs\label{eq:tensor-metric-perturbations}
\beqa
\label{eq:tensor-metric-perturbations-hij}
h^{ij}_\mathbf{k} &=& \frac{1}{\eta}\,
\Big[ \widehat{C}_{\mathbf{k},\,4}\,\sin(k\,\eta)
    + \widehat{C}_{\mathbf{k},\,5}\;\text{sgn}(\eta)\,\cos(k\,\eta) \Big]\,
    e^{ij}_\mathbf{k}\,,
\\[2mm]
\eta &=&
\begin{cases}
+
2\,\sqrt[4]{b^{2}+ t _{0}^{2}}\;\;
 \sqrt[4]{b^{2}+ t^{2}\phantom{ t _{0}^{2}}}\,,
&  \;\;\text{for}\;\;  t  \geq 0\,,
 \\[1mm]
-
2\,\sqrt[4]{b^{2}+ t _{0}^{2}}\;\;
 \sqrt[4]{b^{2}+ t^{2}\phantom{ t _{0}^{2}}}\,,
&  \;\;\text{for}\;\;  t  \leq 0\,,
\end{cases}
\\[2mm]
\eta &\in& (-\infty,\,\widetilde{\eta}_{-}]  \,
\cup \, [\widetilde{\eta}_{+},\,\infty)\,,
\\[2mm]
\widetilde{\eta}_{\pm} &\equiv&
\pm \,2\;\sqrt[4]{b^{2}\,\big(b^{2}+ t _{0}^{2}\big)}
\,,
\eeqa
\esubeqs
with $k\equiv |\mathbf{k}|$ and
constant polarization tensor $e^{ij}_\mathbf{k}$
(the polarization may be different for different wave vectors $\mathbf{k}$).
Remark that, just as in Appendix~\ref{app:Scalar-metric-perturbations},
we have added a sign factor
to the  coefficient $\widehat{C}_{\mathbf{k},\,5}$
in \eqref{eq:tensor-metric-perturbations-hij},
in order to get the proper boundary conditions at
$\eta =\widetilde{\eta}_{\pm}$.

From \eqref{eq:tensor-metric-perturbations-hij}, we see that
short-wavelength gravitational waves ($k\,|\eta| \gg 1$)
have an averaged amplitude that goes as $1/|\eta|$, so that
\beq\label{eq:tensor-metric-perturbations-hij-short-wavelengths}
h^{ij}_\mathbf{k}\Big|^\text{(short-wavelength)}
\sim \frac{\widehat{c}_{\mathbf{k},\,6}}
{\sqrt[4]{1+ t^{2}/b^{2}}}\;e^{ij}_\mathbf{k}\,,
\eeq
for a dimensionless constant $\widehat{c}_{\mathbf{k},\,6}$.
With the cosmic scale factor
$\overline{a}(t) \propto \sqrt[4]{b^{2}+ t^{2}}$,
we observe that the amplitudes
of short-wavelength gravitational waves
from \eqref{eq:tensor-metric-perturbations-hij-short-wavelengths}
go as $1/\overline{a}(t)$, which matches the behavior of the
standard radiation-dominated Friedmann universe; see
the second and third lines below Eq.~(7.100) in Ref.~\cite{Mukhanov2005}.

%%\newpage%%tmp
\section{Imprint from a new phase?}
\label{app:Imprint-from-a-new-phase?}

The present Appendix expands on the discussion of the last paragraph
in Sec.~\ref{sec:Discussion}
by sketching one possible scenario, purely as illustration.

Suppose that a new phase is responsible for the effective spacetime defect
and that scalar metric perturbations (in the Newtonian gauge) emerge
in the corresponding classical spacetime \eqref{eq:newtonian-gauge-scalar-metric}
at cosmic time coordinate $ t =0$.
Matter perturbations in both universes (described by $ t >0$
and $ t <0$) may grow with
thermodynamic time $\mathcal{T}( t ) \equiv | t |$, as discussed in
the last paragraph of Sec.~\ref{sec:Bounce-stability}.

Suppose also that this new phase
sets the value of the metric perturbation $\Phi_\mathbf{k}( t )$
at $ t =0$ differently for comoving wave number
$k \equiv |\mathbf{k}|$ above or below
the critical value $k_\text{crit} \equiv 2\pi/\lambda_\text{crit}$
(the critical physical wave number
is given by $\widetilde{k}_\text{crit} \sim 2\pi/b$),
\bsubeqs\label{eq:Phik0-Phikdot0-from-new-phase}
\beq
\label{eq:Phik0-from-new-phase}
 \Phi_\mathbf{k}(0)=
\begin{cases}
0 \,,   & \text{for}\;\;k   >  k_\text{crit} \,,
 \\[2mm]
1 \,,   & \text{for}\;\;k \leq k_\text{crit} \,.
\end{cases}
\eeq
The new phase is, moreover, supposed to
fix the value of the following quantity involving the
time derivative of the metric perturbation $\Phi_\mathbf{k}( t )$:
\beq
\label{eq:Phikdot0-from-new-phase}
\lim_{ t \to 0}\,
\left|\frac{b^2}{ t }\,\dot{\Phi}_\mathbf{k}( t )\right| \ll 1 \,,
\eeq
\esubeqs
which holds equally for all values of $\mathbf{k}$
and is a stronger condition than $b\,\dot{\Phi}_\mathbf{k}(0)=0$.

From \eqref{eq:Phik0-from-new-phase}
and \eqref{eq:Phikdot0-from-new-phase},
the subsequent metric perturbations
\eqref{eq:amplitude-sol.Phi-matter-case} for $| t | >0$
have the following amplitudes:
\bsubeqs\label{eq:Ck1-Ck2-from-new-phase}
\beqa
\label{eq:Ck1-from-new-phase}
\widehat{C}_{\mathbf{k},\,1} &=&
\begin{cases}
-\widehat{C}_{\mathbf{k},\,2} \,,   & \text{for}\;\;k   >  k_\text{crit} \,,
 \\[2mm]
1-\widehat{C}_{\mathbf{k},\,2} \,,   & \text{for}\;\;k \leq k_\text{crit}\,,
\end{cases}
\\[2mm]
\left|\widehat{C}_{\mathbf{k},\,2}\right|
 &\ll& 1 \,.
\label{eq:Ck2-from-new-phase}
\eeqa
\esubeqs
The implication is that
$\widehat{C}_{\mathbf{k},\,1}$ is
close to $0$ for relatively small wavelengths
and close to $1$ for relatively large wavelengths,
as long as the assumptions hold true.
From \eqref{eq:Ck1-Ck2-from-new-phase},
the spectrum of adiabatic density perturbations \eqref{eq:sol.density-perturb.}
then shifts
as $k$ changes from above $k_\text{crit}$ to below $k_\text{crit}$,
with a jump of the coefficient multiplying the growing
mode proportional to $\big(b^{2}+ t^{2}\big)^{1/3}$.

It is an interesting problem to see if it is possible to match
the perturbations of the degenerate metric to the results from
loop quantum gravity~\cite{Ashtekar2002},
string theory~\cite{Witten2002},
or emergent gravity~\cite{Laughlin2003}.
For homogeneous models, Appendix~B of Ref.~\cite{Klinkhamer2019-More}
has already given a first comparison between
extended general relativity with an appropriate degenerate metric
and the effective theories from loop quantum cosmology
and string cosmology.

\end{appendix}

\newpage%%tmp

\end{document}